\documentclass{cernrep}

\usepackage{wrapfig}

\pdfoutput=1

\begin{document}
\title{The photon PDF from high-mass Drell-Yan data at the LHC}
\author{V. Bertone\thanks {On behalf of the {\tt xFitter} developer's
    team.}}

\institute{Department of Physics and Astronomy, VU University, NL-1081
  HV Amsterdam, \\and Nikhef Theory Group Science Park 105, 1098 XG
  Amsterdam, The Netherlands}

\begin{abstract}
  I present a determination of the photon PDF from a fit to the recent
  ATLAS measurements of high-mass Drell-Yan lepton-pair production at
  $\sqrt{s} = 8$ TeV.  This analysis is based on the {\tt xFitter}
  framework interfaced to the {\tt APFEL} program, that accounts for
  NLO QED effects, and to the {\tt aMCfast} code to account for the
  photon-initiated contributions within {\tt MadGraph5\_aMC@NLO}. The
  result is compared with other recent determinations of the photon
  PDF finding a general good agreement. This contribution is based on
  the results presented in Ref.~\cite{Giuli:2017oii}.
\end{abstract}


\maketitle

\textit{\small PHOTON'17  Conference Proceedings, CERN 22 - 27 May
  2017, to appear on CERN Report}

\section{Introduction and motivation}

In order to achieve accurate predictions for the LHC phenomenology,
QCD corrections, where NNLO is becoming the standard, have to be
supplemented with electroweak (EW) effects. One of the direct
consequences of these corrections is the introduction of the photon
PDF.

An number of determinations of the photon PDF based on a variety of
different approaches has been achieved in the
past~\cite{Martin:2004dh,Ball:2013hta,Schmidt:2015zda,Bertone:2016ume,Ball:2014uwa,Manohar:2016nzj,Harland-Lang:2016kog}.
The aim of this particular work is to obtain a model-independent
determination of the photon PDF exploiting the recent high-mass
Drell-Yan measurements at $\sqrt{s}=8$ TeV from
ATLAS~\cite{Aad:2016zzw}, that have proven to provide a significant
constraint on this distribution.

The constraining power of the Drell-Yan process on the photon PDF can
be easily understood in terms of Feynman diagrams. Indeed, in the
presence of EW corrections, the partonic channel
$\gamma\gamma\rightarrow \ell^+\ell^-$ contributes to the leading
order (LO) cross section for lepton-pair production in $pp$
scattering.
\begin{figure}[h]
  \begin{center}
    \includegraphics[width=0.8\textwidth]{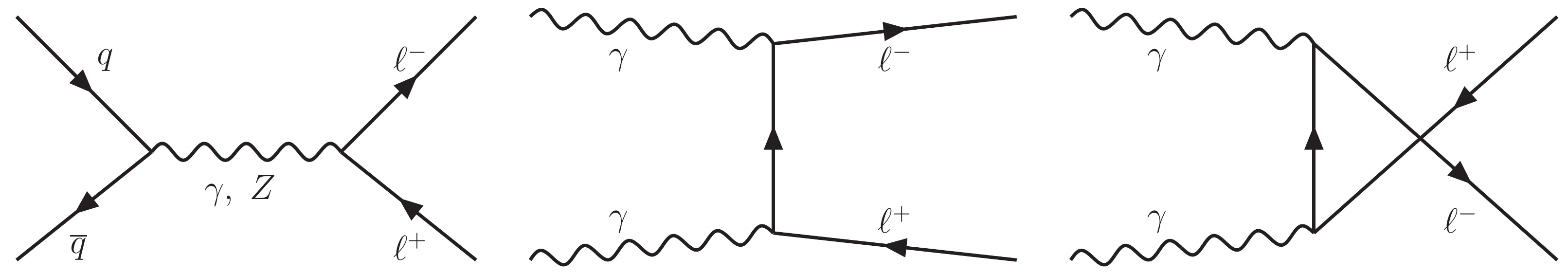}
  \end{center}
  \caption{\footnotesize LO diagrams that contribute to lepton-pair
    production at hadron colliders.}
  \label{fig:LODY}
\end{figure}
This is illustrated in Fig.~\ref{fig:LODY} where the LO diagrams
contributing to this process are shown.

The high invariant-mass distribution of the lepton pair is
particularly relevant because this observable is such that the
$\gamma\gamma$ contribution, despite the relatively small size of the
photon PDF, becomes comparable to that induced by the $q\overline{q}$
channel.

\begin{wrapfigure}{r}{0.5\textwidth}
 \vspace{-20pt}
 \begin{center}
   \includegraphics[width=0.48\textwidth]{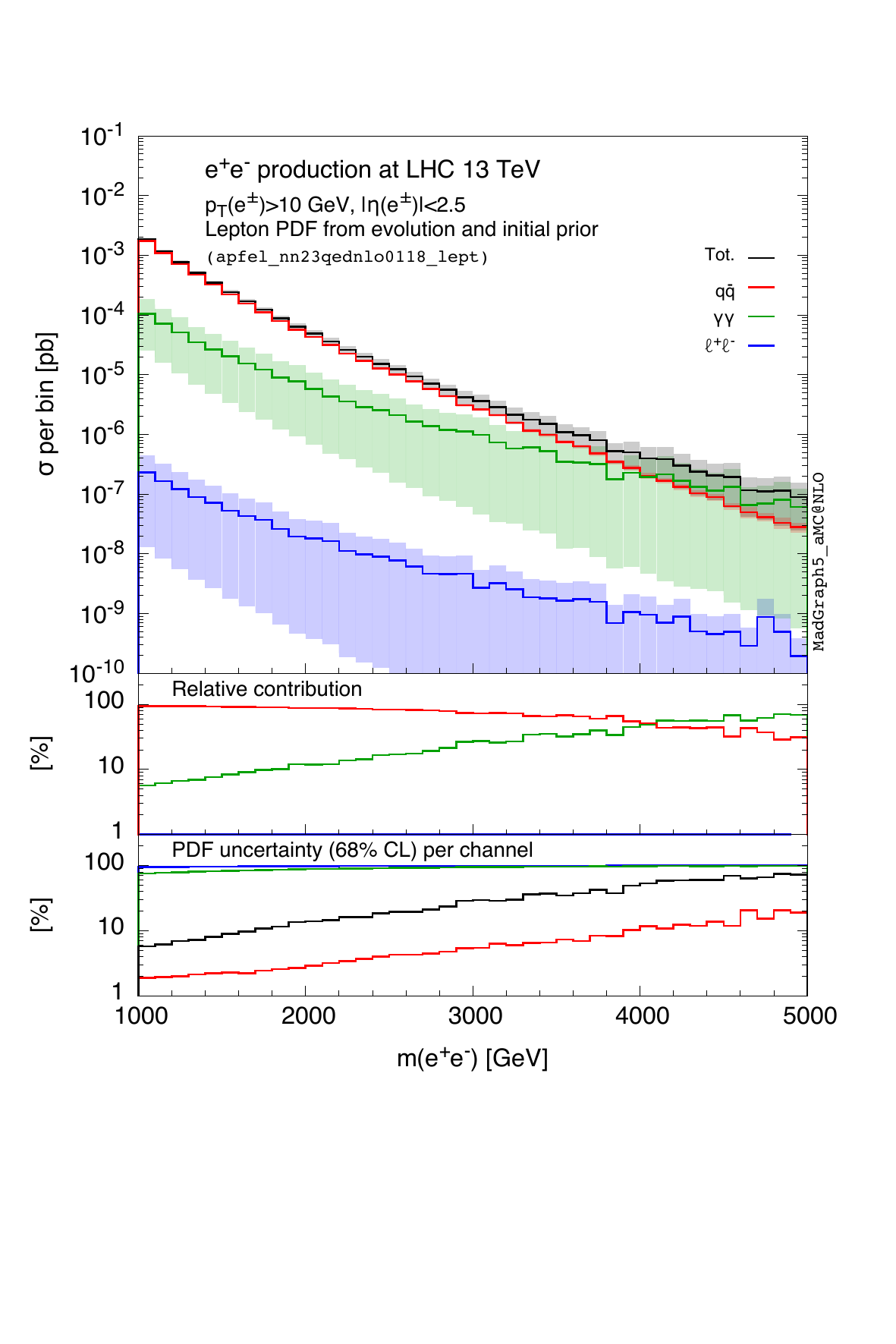}
 \end{center}
 \vspace{-70pt}
 \caption{\footnotesize Predictions at LO for the lepton-pair
   invariant mass distribution in $e^+e^−$ production at the LHC at 13
   TeV~\cite{Bertone:2015lqa}.}
 \label{fig:HMDY}
\end{wrapfigure}
\noindent As an illustration, the LO prediction for the lepton-pair
invariant mass distribution in $e^+e^−$ production at the 13 TeV LHC
is shown in Fig.~\ref{fig:HMDY}~\cite{Bertone:2015lqa}. This plot
indicates that the $\gamma\gamma$ channel becomes increasingly
important at large values of the invariant mass and eventually
dominates the distribution. Based on simple kinematic considerations,
one can show that the high invariant-mass distribution in lepton-pair
production probes the photon PDF at relatively large values of
Bjorken-$x$, indicatively $x\gtrsim 0.02$.

This observation constitutes a compelling motivation to exploit the
precise experimental data produced by the LHC, such as the recent
\mbox{ATLAS} data at 8 TeV published in Ref.~\cite{Aad:2016zzw}, to
constrain the photon PDF in this region.

A crucial aspect of this analysis is the consistent inclusion of the
relevant EW corrections. As it was shown in
Ref.~\cite{Boughezal:2013cwa}, the Drell-Yan process receives sizeable
pure weak corrections that almost balance the corrections induced by
the photon-initiated channels. Therefore, the inclusion of the NLO EW
corrections to the computation of the Drell-Yan cross sections is
extremely important to achieve a reliable determination of the photon
PDF.

This study was carried out within the open-source {\tt xFitter}
framework~\cite{Alekhin:2014irh} that provides a unique environment to
extract PDFs from experimental data.

\section{The dataset}\label{sec:dataset}

As mentioned above, our determination of the photon PDF relies on the
recent ATLAS 8 TeV high-mass Drell-Yan
data~\cite{Aad:2016zzw}. Measurements are delivered in three different
formats:
\begin{enumerate}
\item as single-differential cross-section distributions in the
  lepton-pair invariant mass $m_{ll}$,
\item as double-differential cross-section distributions in $m_{ll}$
  and the rapidity of the lepton pair $|y_{ll}|$,
\item and as double-differential cross-section distributions in
  $m_{ll}$ and the difference in pseudo-rapidity between the two
  leptons $\Delta\eta_{ll}$.
\end{enumerate}
In our analysis we have chosen to use the second format that counts 48
data points distributed in 5 $m_{ll}$ bins: [116-150], [150-200],
[200-300], [300-500], [500-1500] GeV. The first three (last two)
$m_{ll}$ bins are divided into 12 (6) bins in $|y_{ll}|$ extending up
to 2.4. The relevant analysis cuts on the data are: $m_{ll}\geq 116$
GeV, $|\eta_{ll}|\leq 2.5$, and $p_T^l\geq 40$ GeV (30) GeV for the
leading (sub-leading) lepton.

The ATLAS data alone would clearly be insufficient to carry out an
analysis aimed at the extraction of a reliable set of PDFs. Therefore,
this data is supplemented by the combined inclusive deep-inelastic
scattering (DIS) cross-section data from
HERA~\cite{Abramowicz:2015mha}, on which we imposed the cut
$Q^2\geq Q_{\rm min}^2 = 7.5$ GeV$^2$. While the ATLAS data is
directly sensitive to the photon PDF, the HERA data carries detailed
information on the quark and gluon content of the proton.  The union
of these two datasets allows us to perform a solid determination of a
the proton PDFs.

\section{Electroweak corrections}\label{sec:theory}

A central aspect of this analysis is the inclusion of the EW
effects. More in particular, we employed predictions accurate to NNLO
in QCD and consistently included NLO EW/QED corrections. In the
present analysis, this concerns three main sectors which I discuss in
turn: the QED corrections to the evolution of PDFs, the QED
corrections to the DIS structure functions, and the full EW
corrections to the Drell-Yan cross sections.

\subsection{Evolution}

The evolution of PDFs is governed by the DGLAP equations. The DGLAP
splitting functions are known up to $\mathcal{O}(\alpha_s^3)$, $i.e.$
NNLO in QCD, since long~\cite{Moch:2004pa,Vogt:2004mw}. The QED
corrections are instead much more recent. The $\mathcal{O}(\alpha)$
corrections, where $\alpha$ is the QED coupling, were derived in
Ref.~\cite{Martin:2004dh}, while the $\mathcal{O}(\alpha\alpha_s)$ and
$\mathcal{O}(\alpha^2)$, which represent the NLO QED corrections,
where computed in Refs.~\cite{deFlorian:2015ujt,deFlorian:2016gvk}.

The implementation of the full NLO QED corrections to the DGLAP
evolution was achieved very recently in the {\tt APFEL}
program~\cite{Bertone:2013vaa} following the approach of
Ref.~\cite{Bertone:2015lqa} and documented in
Ref.~\cite{Giuli:2017oii}. A cross-check of the implementation was
performed using the independent {\tt QEDEVOL}
code~\cite{Sadykov:2014aua} based on the {\tt QCDNUM} evolution
program~\cite{Botje:2010ay}.

\begin{wrapfigure}{r}{0.5\textwidth}
\vspace{-40pt}
 \begin{center}
\includegraphics[width=0.55\textwidth]{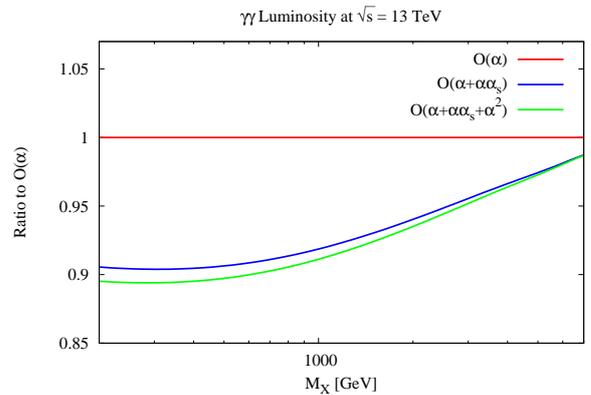}
 \end{center}
\vspace{-40pt}
\caption{\footnotesize The $\gamma\gamma$ luminosity at
  $\sqrt{s} = 13$ TeV as a function of the final state invariant mass
  $M_X$. The curves, taking into account the $\mathcal{O}(\alpha)$,
  the $\mathcal{O}(\alpha+\alpha_s\alpha)$, and the complete
  $\mathcal{O}(\alpha+\alpha_s\alpha+\alpha^2)$ effects in the DGLAP
  evolution, are presented as ratios to the $\mathcal{O}(\alpha)$
  result. The calculation was performed using NNPDF3.0QED NNLO
  set.}\label{fig:GammaGammaLumi}
\end{wrapfigure}
The effect of the NLO QED corrections on the $\gamma\gamma$
luminosity, relevant to the computation of the Drell-Yan invariant
mass distribution, is shown in Fig.~\ref{fig:GammaGammaLumi} as a
function of the final state invariant mass $M_X$. The photon PDF taken
from the NNPDF3.0QED set is evolved including in the DGLAP equation
the $\mathcal{O}(\alpha)$, the $\mathcal{O}(\alpha+\alpha_s\alpha)$,
and the complete $\mathcal{O}(\alpha+\alpha_s\alpha+\alpha^2)$
corrections. Results are shown as ratios to the $\mathcal{O}(\alpha)$
curve. While the effect of the $\mathcal{O}(\alpha^2)$ is very mild,
the impact of the $\mathcal{O}(\alpha_s\alpha)$ at relatively small
values of $M_X$ can be as big as 10\%. This reduces to around 3-5\% at
large invariant masses: this is the region of relevance in our study.
This is still a significant effect due to the experimental uncertainty
of the ATLAS data and thus it is important to take it into account.

In principle, the NLO QED corrections influence also the running of
the QCD coupling $\alpha_s$ and the QED coupling $\alpha$. In fact,
they introduce additional mixing terms in the respective
$\beta$-functions that couple the evolution of the two
couplings. However, it turns out that the impact of the mixing terms
is tiny on both couplings and thus we decided not to included them as
this would uselessly complicated the implementation.

\subsection{DIS structure functions}

When considering NLO QED corrections to the DIS structure functions,
it is necessary to include into the hard cross sections all the
$\mathcal{O}(\alpha)$ diagrams. The coefficient functions of these
diagrams, being of purely QED origin, can be easily derived from the
QCD expressions by properly adjusting the colour factors. This
correspondence holds irrespective of whether mass effects are
included. This allowed for an easy implementation of the QED
corrections to the FONLL general-mass
scheme~\cite{Forte:2010ta}. Specifically, in this work we have used
the variant C of the FONLL scheme, accurate to NNLO in QCD,
supplemented by the NLO QED corrections.

\begin{wrapfigure}{r}{0.5\textwidth}
\vspace{-40pt}
 \begin{center}
   \includegraphics[width=0.55\textwidth]{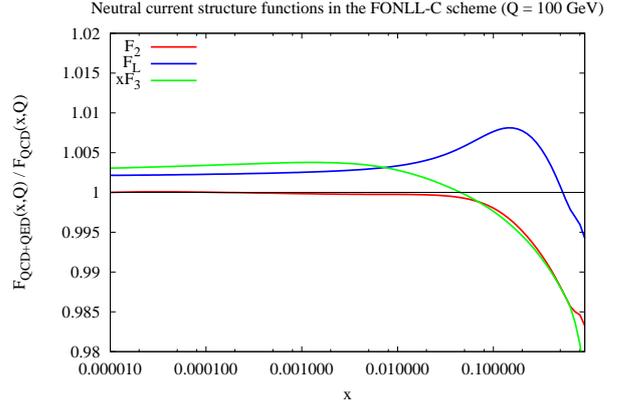}
\end{center}
\vspace{-40pt}
\caption{\footnotesize The effects of the NLO QED corrections on the
  neutral-current DIS structure functions $F_2, F_L$ and $xF_3$ at
  $Q=100$ as functions of $x$, normalised to the pure QCD results. The
  calculation has been performed in the FONLL-C scheme using
  NNPDF3.0QED NNLO evolved including the QED corrections to the
  evolution discussed above.}\label{fig:StructFuncs}
\end{wrapfigure}
An intereresting feature of the NLO QED corrections to the DIS 
structure function is that they introduce photon-initiated diagrams 
providing a direct handle on the photon PDF. 

Fig.~\ref{fig:StructFuncs} displays the effect of the NLO QED
corrections on the neutral-current structure functions $F_2$, $F_L$,
and $xF_3$. The predictions have been obtained including the NLO QED
corrections also to the DGLAP evolution and are shown normalised to
the pure QCD results. The impact of the NLO QED corrections is very
moderate especially at low $x$ but becomes more significant at large
$x$, where it is of the order of 2\%. The same behaviour is observed
also for the charged-current structure functions

Although the net effect of the NLO QED corrections on the DIS
structure functions is small, it is significant when compared to the
ty\-pi\-cal size of the uncertainties of the HERA combined data. This
implies that the DIS data, despite very moderately, contributes to
constrain the photon PDF in the large-$x$ region.

\subsection{Drell-Yan cross sections}

For the calculation of the Drell-Yan cross sections at NLO in QCD, we
have used the {\tt
  MadGraph5}\-{\tt{\_}aMC@\-NLO}~\cite{Alwall:2014hca} program
interfaced to {\tt APPLgrid}~\cite{Carli:2010rw} through {\tt
  aMCfast}~\cite{Bertone:2014zva}. The computation also includes the
contribution from the photon-initiated diagrams shown in
Fig.~\ref{fig:LODY}. Finite mass effects of charm and bottom quarks in
the matrix elements are neglected, as appropriate for a high-scale
process.

The NLO calculations are supplemented by $K$-factors obtained with the
{\tt FEWZ} code~\cite{Gavin:2012sy} to account for the NNLO QCD and
the NLO EW corrections. The $K$-factors are defined as:
\begin{equation} \label{eq:kfactor}
  K(m_{ll},|y_{ll}|) \equiv\frac{\rm NNLO\  QCD  + NLO\  EW}{\rm NLO\  QCD + LO\  EW} \, ,
\end{equation}
\begin{wrapfigure}{r}{0.5\textwidth}
  \vspace{-25pt}
 \begin{center}
   \includegraphics[width=0.55\textwidth]{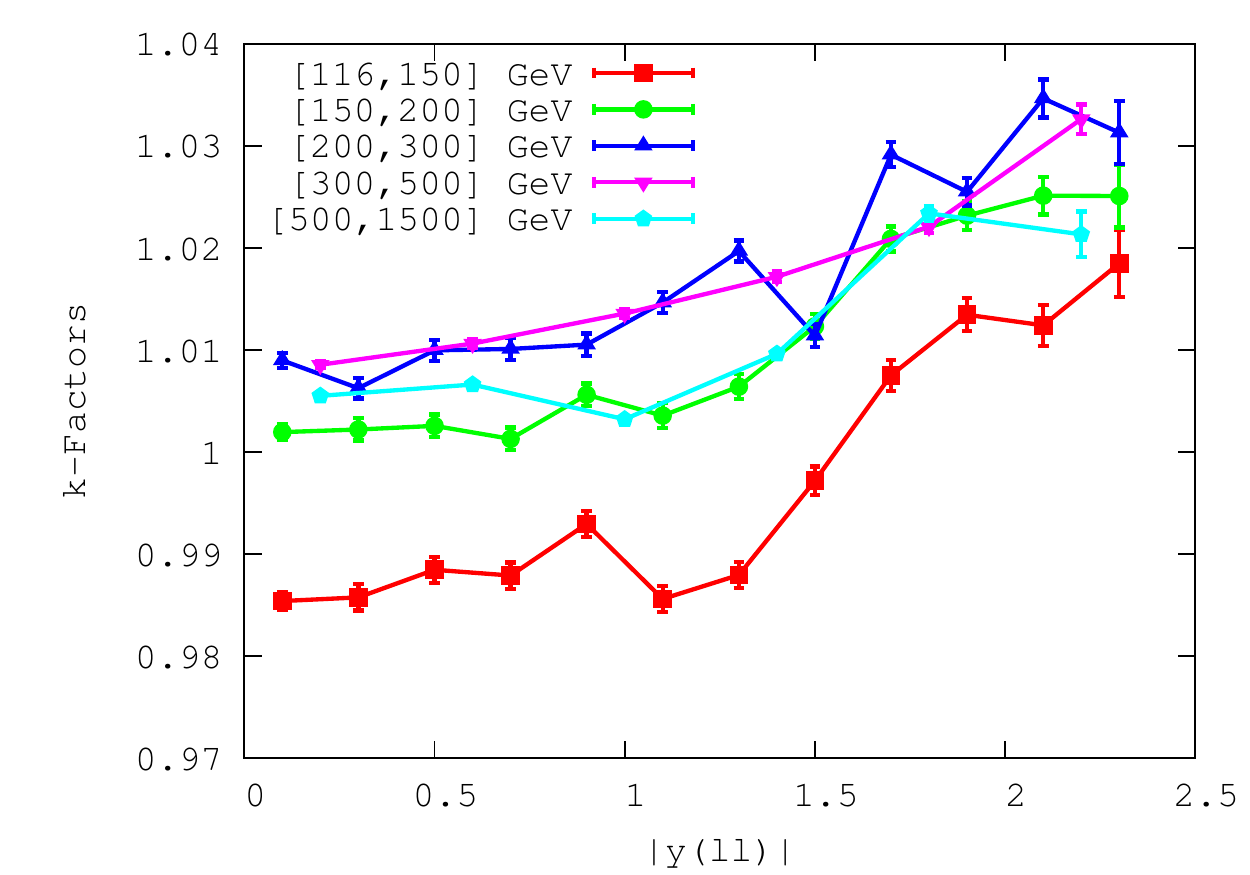}
\end{center}
  \vspace{-20pt}
\caption{\footnotesize The $K$-factors, defined in
  Eq.~(\ref{eq:kfactor}), as a function of the lepton-pair rapidity
  $|y_{ll}|$ for each $m_{ll}$ bin.}
\label{fig:kf}
\end{wrapfigure}
and computed using the MMHT2014 NNLO~\cite{Harland-Lang:2014zoa} PDF 
set both in the numerator and in the denominator. 

Fig.~\ref{fig:kf} shows the $K$-factors of Eq.~(\ref{eq:kfactor}) as a
function of the lepton-pair rapidity $|y_{ll}|$ for each $m_{ll}$
bin. The points correspond to the kinematics of double-differential
distributions in $(m_{ll},|y_{ll}|)$ of the ATLAS high-mass Drell-Yan
data included in our analysis.
 
The $K$-factors vary between 0.98 and 1.04, indicating that
higher-order corrections are ge\-ne\-ral\-ly moderate. The trend
follows the expectation: the $K$-factors are particularly small at low
invariant masses and in the central region, and tend to grow at larger
values of $m_{ll}$ and in the forward region where they can be as
large ar 4\%.

\newpage
\section{Fit settings}

Our determination of the photon PDF, along with quark and gluon PDFs,
was carried out in the {\tt xFitter} framework interfaced to the {\tt
  APFEL} code. The dataset included in our fit was discussed in
Sect.~\ref{sec:dataset} and the theory setup presented in
Sect.~\ref{sec:theory}. In this section we discuss the fit settings.

We parametrise the following six independent distributions at the
initial scale $Q_0$:
\begin{equation}\label{eq:parametrization}
\begin{array}{rcl}
  xu(x,Q_0)-x\overline{u}(x,Q_0)\equiv xu_v(x,Q_0) &=& A_{u_v}x^{B_{u_v}}(1-x)^{C_{u_v}}(1+E_{u_v}x^{2})\, , \\
  xd(x,Q_0)-x\overline{d}(x,Q_0)\equiv xd_v(x,Q_0) &=& A_{d_v}x^{B_{d_v}}(1-x)^{C_{d_v}}\, , \\
  x\overline{u}(x,Q_0)\equiv x\bar{U}(x,Q_0) &=& A_{\bar{U}}x^{B_{\bar{U}}}(1-x)^{C_{\bar{U}}}\, , \\
  x\overline{d}(x,Q_0)+x\overline{s}(x,Q_0)\equiv x\bar{D}(x,Q_0) &=& A_{\bar{D}}x^{B_{\bar{D}}}(1-x)^{C_{\bar{D}}}\, , \\
  xg(x,Q_0) &=& A_{g}x^{B_{g}}(1-x)^{C_{g}}(1+E_{g}x^{2})\, , \\
  x\gamma(x,Q_0) &=& A_{\gamma}x^{B_{\gamma}}(1-x)^{C_{\gamma}}(1+D_{\gamma}x+E_{\gamma}x^{2}) \,.
\end{array}
\end{equation}
The parameters $B_{\bar{U}}$ and $B_{\bar{D}}$ are set equal so that
the quark sea distributions have the same small-$x$ behaviour.
Moreover, we assume $x\bar{s} = r_s x\bar{d}$, with
$r_s = 1$~\cite{Aaboud:2016btc}, and $A_{\bar{U}} = A_{\bar{D}} / 2$,
such that $x\bar{u}\to x\bar{d}$ for $x \to 0$.

The numerical values of the heavy-quark masses are taken to be
$m_c=1.47$~GeV and $m_b=4.5$~GeV. The reference values of the QCD and
QED couplings are chosen to be $\alpha_s(M_Z)=0.118$ and
$\alpha(m_\tau=1.777\mbox{ GeV})=1/133.4$. As for the initial scale,
we choose $Q_0>m_c =\sqrt{7.5}$~GeV such that it is below the scale of
all data points included in our fit. This particular value of the
initial scale $Q_0$ is peculiar as compared to the typical choice
$Q_0< m_c\simeq 1$~ GeV. The reason for choosing a somewhat larger
scale is that it helps stabilise the photon PDF. However, in order to
still be able to generate the charm PDFs perturbatively without the
need to parameterise them, we exploited one of the functionalities of
{\tt APFEL} to set charm threshold
$\mu_c = Q_0 > m_c$~\cite{Bertone:2017ehk}.

\section{Results}

\begin{wraptable}{r}{0.55\textwidth}
  \footnotesize
  \vspace{-10pt}
  \centering
  \begin{tabular}{|c|c|}
    \hline
    Dataset  &   $\chi^2$ /$N_{\rm dat}$ \\
    \hline
    \hline
    HERA combined DIS data & 1236/1056\\
    \hline
    ATLAS DY data [116 GeV $\le m_{ll} \le $ 150 GeV]  &  9/12 \\
    ATLAS DY data [150 GeV $\le m_{ll} \le $ 200 GeV]  &  15/12 \\
    ATLAS DY data [200 GeV $\le m_{ll} \le $ 300 GeV]  &  14/12 \\
    ATLAS DY data [300 GeV $\le m_{ll} \le $ 500 GeV]  &  5/6 \\
    ATLAS DY data [500 GeV $\le m_{ll} \le $ 1500 GeV] &  4/6 \\
    \hline
    Total  ATLAS DY data $\chi^2/N_{\rm dat}$  & 48/48 \\
    \hline
    \hline
    Combined HERA I+II and high-mass DY $\chi^2/N_{\rm dof}$   & 1284/1083 \\
    \hline
    \end{tabular}
    \caption{\footnotesize The $\chi^{2}/N_{\rm dat}$ of fit to the
      combined HERA data and to the five $m_{ll}$ bins of the ATLAS
      Drell-Yan data.  The global $\chi^2/N_{\rm dof}$ is also
      reported, where $N_{\rm dof}$ is the number of degrees of
      freedom in the fit.}
    \label{tab:chi2fit}
\end{wraptable}
I finally turn to discuss the results of our fit. The partial
$\chi^2$'s normalised to the number of data points for the HERA
combined data and for the ATLAS high-mass Drell-yan data (bin by bin
in $m_{ll}$ and total), as well as the total $\chi^2$ normalised to
the number of degrees of freedom, are reported in
Tab.~\ref{tab:chi2fit}.

The overall fit quality is acceptably good. On the one hand, the
description of the HERA data is comparable to that achieved in the
HERAPDF2.0 analysis~\cite{Abramowicz:2015mha}. On the other hand,
despite the small experimental uncertainties, the ATLAS Drell-Yan data
is perfectly fitted with a $\chi^2/N_{\rm dat}$ equal to
$48/48$. Remarkably, the single $m_{ll}$ bins of this dataset have all
a good $\chi^2$.

\begin{figure}[h]
\begin{center}
  \includegraphics[width=6cm]{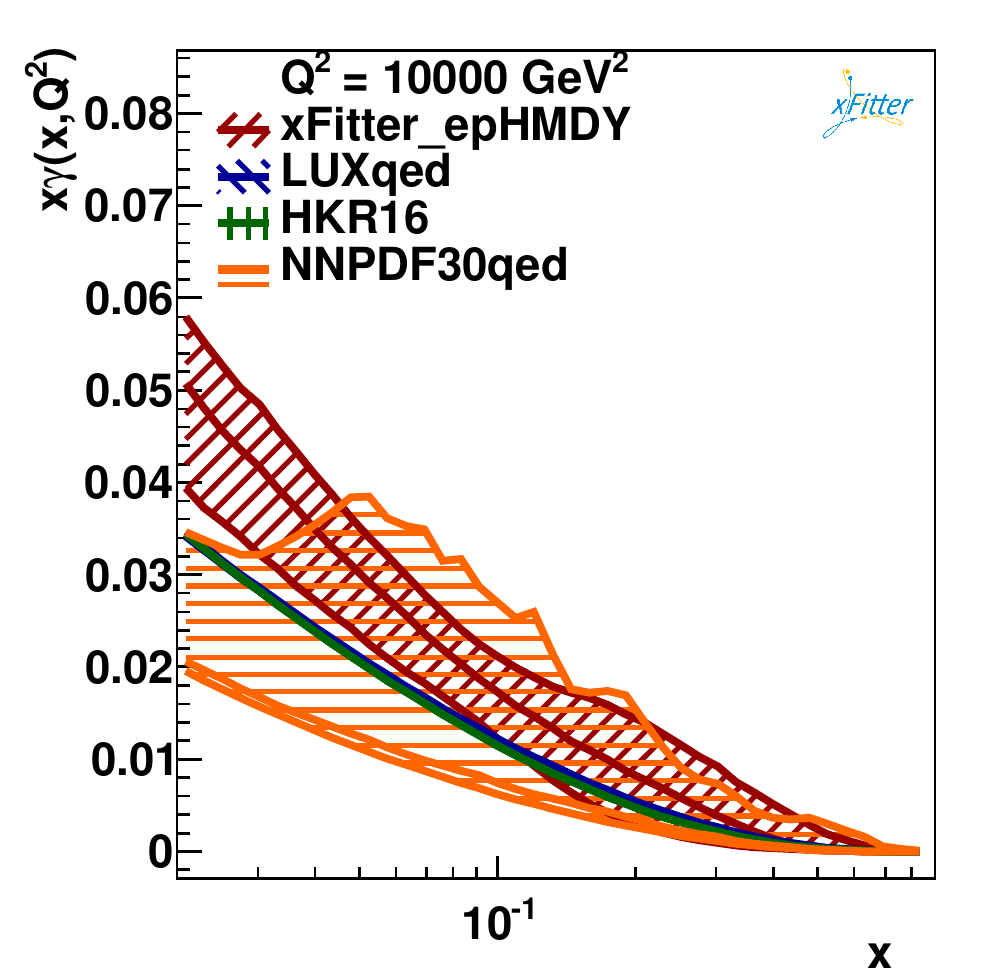}
  \includegraphics[width=6cm]{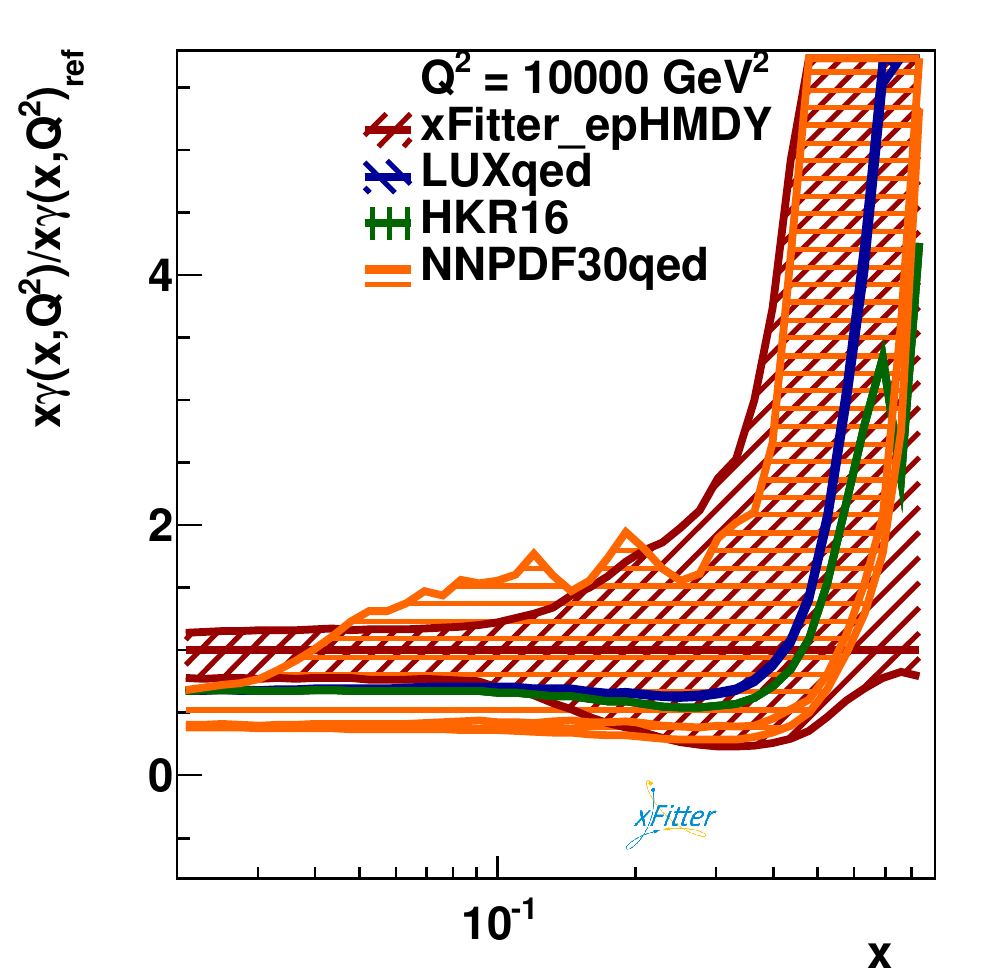}
\end{center}
\caption{\footnotesize Left plot: comparison between the photon PDF at
  $Q^2=10^4$ GeV$^2$ from the present analysis ({\tt xFitter\_epHMDY})
  and the results from LUXqed, HKR16, and NNPDF3.0QED. Right plot:
  same as the left plot but with the distributions normalised to the
  central value of {\tt xFitter\_epHMDY}. The uncertainty bands
  represent the 68\% confidence level. For HKR16 only the central
  value is available.}
  \label{fig:photon}
\end{figure}
The photon PDF at $Q^2=10^4$ GeV$^2$ obtained from our analysis, that
we dubbed {\tt xFitter\_epHMDY}, is shown in Fig.~\ref{fig:photon} and
compared to the LUXqed~\cite{Manohar:2016nzj}, the
HKR16~\cite{Harland-Lang:2016kog}, and the
NNPDF3.0QED~\cite{Bertone:2016ume} results. The absolute distributions
are shown in the lef plot, while they are displayed as ratios to the
central value of {\tt xFitter\_epHMDY} in the right plot. The
uncertainty bands represent the 68\% confidence level for all
distributions but for HKR16 for which only the central value is made
available by the authors. The $x$-range shown Fig.~\ref{fig:photon} is
limited to the region $0.02 \le x \le 0.9$ where the ATLAS Drell-Yan
data are expected to constrain the photon PDF.

Fig.~\ref{fig:photon} shows that for $x\ge 0.1$ the four
determinations are consistent within PDF uncertainties. For smaller
values of $x$, the photon PDFs from LUXqed and HKR16 are lower than
{\tt xFitter\_epHMDY} but the agreement remains at the 2-$\sigma$
level. A better agreement with the NNPDF3.0QED photon PDF is observed
all over the considered range also due to the larger
uncertainties. Interestingly, Fig.~\ref{fig:photon} also shows that
for $0.04 \le x \le 0.2$ the present analysis exhibits smaller PDF
uncertainties as compared to those of NNPDF3.0QED. We conclude that
this is the effect of the constraining power of the ATLAS Drell-Yan
data used in this analysis but not in NNPDF3.0QED. We also observe
that this dataset has very little impact of the other PDFs.

Finally, in order to assess the robustness of our fit, we have
performed a number of variations with respect to the default settings.
Specifically, we considered variations of: the values of the input
physical parameters, such as $\alpha_s$, the heavy-quark masses $m_c$
and $m_b$, and the strangeness fraction $r_r$; the PDF parametrisation
and the input scale $Q_0$; the cut $Q_{\rm min}^2$ on the scale of the
data included in the fit. In all cases, the resulting distributions of
a given variation were in agreement, typically well within one
standard deviation, with the result obtained with the default
settings.

The {\tt xFitter\_epHMDY} presented in this work is available in the
{\tt LHAPDF6} format~\cite{Buckley:2014ana} upon request from the
authors.

\section*{Acknowledgements}

I would like to heartily thank Ringaile Pla{\v c}akyt{\.e} and Voica
Radescu for their invaluable contribution to this work and for their
outstanding dedication as conveners of the {\tt HERAFitter-xFitter}
project from 2012 until May 2017.

\end{document}